\pdfoutput=1 

\documentclass[manuscript]{aastex}
\usepackage{ulem}
\usepackage{natbib}

\begin{document}

\title{The Illumination and Growth of CRL\,2688: \\ An Analysis of New \& Archival HST Observations}
\author{Bruce Balick$^1$, Thomas Gomez$^1$, Dejan Vinkovi\'{c}$^2$,\\  Javier Alcolea$^3$, Romano L.M. Corradi$^{4,5}$, Adam Frank$^6$}

\affil{$^1$Department of Astronomy, University of Washington, Seattle, WA 98195-1580, USA}
\affil{$^2$Physics Department, University of Split, Teslina 12/III, HR-21000 Split, Croatia}
\affil{$^3$Observatorio Astron$\acute{o}$mico Nacional (IGN), E-28014 Madrid, Spain}
\affil{$^4$Instituto de Astrof{\'{\i}}sica de Canarias, E-38200 La Laguna, Tenerife, Spain}
\affil{$^5$Departamento de Astrof{\'{\i}}sica, Universidad de La Laguna, E-38206 La Laguna, Tenerife, Spain}
\affil{$^6$Department of Physics and Astronomy, University of Rochester, Rochester, NY 14627, USA}
\email{balick@astro.washington.edu, gomezt@u.washington.edu, j.alcolea@oan.es,  vinkovic@pmfst.hr, rcorradi@iac.es, afrank@pas.rochester.edu}

\begin{abstract}
We present four-color images of CRL\,2688 obtained in 2009 using the Wide-Field Camera 3 on HST. The F606W image is compared with archival images in very similar filters to monitor the proper motions of nebular structure. We find that the bright N--S lobes have expanded uniformly by 2.5\% and that the ensemble of rings has translated radially by $0\farcs07$ in 6.65 y. The rings were ejected every 100y for $\sim$4 millennia until the lobes formed 250y ago. Starlight scattered from the edges of the dark E--W dust lane is coincident with extant H$_2$ images and leading tips of eight pairs of CO outflows. We interpret this as evidence that fingers lie within geometrically opposite cones of opening angles $\approx30^\circ$ like those in CRL618. By combining our results of the rings with $^{12}$CO absorption from the extended AGB wind we ascertain that the rings were ejected at $\sim$18 km s$^{-1}$ with very little variation and that the distance to CRL\,2688, $v_{exp}/\dot\theta_{exp}$, is 300-350 pc. Our 2009 imaging program included filters that span 0.6 to 1.6$\micron$. We constructed a two-dimensional dust scattering model of stellar radiation through CRL\,2688 that successfully reproduces the details of the nebular geometry, its integrated spectral energy distribution, and nearly all of its color variations. The model implies that the optical opacity of the lobes $\ga1$, the dust particle density in the rings decreases as radius$^{-3}$ and that the mass and momentum of the AGB winds and their rings have increased over time.
\end{abstract}

\keywords{stars: AGB and post-AGB, stars: winds, outflows, (ISM:) planetary nebulae: individual (CRL\,2688), (ISM:) reflection nebulae}

\section{Introduction}

CRL\,2688\footnote{ aka Egg Nebula, RAFGL2688, V* V1610 Cyg; RA$_{2000}$ = 21 02 18.8, $\delta_{2000}$ = +36 41 41, l,b = 080.2-06.5} is the ideal pre-planetary nebula (``pPN'') for studying the historical ejection of mass from many AGB and post-AGB stars.  Optically, its bright $\sim15\arcsec$ nebular core contains a polar (N--S) pair of bright expanding lobes and a dark equatorial (E--W) dust lane, both of which are presumed to have formed at about the same time as the transition from the AGB to post-AGB states of late stellar evolution. These features are characteristic of pPNe in the DUPLEX subclass (functionally, those that contain deep equatorial dust extinction) (\citealt{2000ApJ...528..861U, 2011AJ....141..134S}\footnote{  see also the compilation of images of pPNe at \url{www.astro.washington.edu/balick/pPNe/}}.  In addition an ensemble of concentric, quasi-circular arcs (hereafter ``rings'')  of total radial extent $\ga40\arcsec$ surrounds the lobes and the dust lane.  The arcs are spaced semi regularly with characteristic separations  $\sim1\arcsec$.   The high brightness of the lobes and the rings as well as their proximity to make CRL\,2688 a particular productive site to study their characteristics.

CRL\,2688 has a wealth of complementary observational information.  For example, the visible and near-IR illumination of CRL\,2688 consists of direct and scattered light from a carbon- and nitrogen-rich but completely obscured supergiant of apparent spectral type F5 Ia with T$_{\rm{eff}}$ $\approx$ 6500\,K \citep{2000AstL...26..439K, 1977PASP...89..829C,1975ApJ...198L.135C}.  Lying adjacent to the nebular center of symmetry are smaller structures seen in the radio continuum \citep{2000ApJ...528L.105J}, the mid infrared \citep[hereafter ``G02'']{2007apn4.confE..55L, 2002ApJ...572..276G}, and a cavity of diameter $2\arcsec$ seen in $\rm^{12}$CO (2--1) \citep[hereafter ``C00'']{2000A&A...353L..25C}.  Like the star, these regions are fully obscured within the dust lane at near IR and visible wavelengths. (We largely ignore them in this paper since they do not appear to affect the structure of HST images.)  Further afield, the bases of both of the lobes are especially prominent in the thermal infrared (G02), suggesting that starlight scatters into the lobes through dense and warmed dust that might affect its initial colors.   Finally, H$_2$ emission outlines the perimeter of the dust lane and the tips and edges of the N--S lobes \citep[hereafter ``S98a'']{2007apn4.confE..55L, 2001ApJ...546..279K, 1998ApJ...492L.163S}. The shocked H$_2$ traces the interaction zones between the (presumably) expanding lobes and dust lane, on the one hand, and the ambient halo containing the rings, on the other.  

Lying outside of the lobes and the dust lane, the well illuminated, delineated, and resolved rings of CRL\,2688 represent the early history of AGB mass ejection.   \citet{2001AJ....121.2775H} found similar but less conspicuous rings in IRAS16594-04656 and IRAS20028+3910.  Other rings have been seen in IRAS 17441-2411 by \citet{1998ApJ...508..744S}, IRAS17150-3224 by \citet{1998ApJ...501L.117K}, IRAS19114+0002  by Su 2003\footnote{  
\url{www.astro.washington.edu/users/balick/APN/APN\_tmeeting\_tdocuments/Tuesday/ksu.ppt}}, 
IRC+10216 by \citet{2000A&A...359..707M}.  Of these, the rings of CRL\,2688 are the best illuminated and defined.  However, aside from their density modulations little is known of the ages, expansion patterns, and ejection histories of any of these ensembles of rings.  

The broad focus in this paper is the evolution patterns of the lobes and the rings.  Thus we briefly review observations of their kinematics here. So far the only single-dish CO studies can detect the flux from the entire ensemble of the rings (e.g., \citealt[hereafter B01]{2001A&A...377..868B}, \& \citealt{2005ApJ...624..331H}).  Obviously such studies cannot resolve the expansion pattern.   Spatially resolved kinematic studies of CRL\,2688 consist primarily of $^{\rm{12}}$CO observations. \citet[hereafter ``C00'']{2000A&A...353L..25C} found several pairs of fingers confined  within the lobes and the dust lane. (C00 elected to associate each of the fingers with collimated ``jets''.)  However, they could not detect the low-brightness halo with the rings in their study.  A few years later \citet{2006ApJ...652.1626F} observed $^{\rm{12}}$CO in CRL\,2688 using closer antenna spacings and better sensitivity to large structures. They marginally detected the inner portions of the halo containing the rings, but could not fit an expansion pattern to this zone.  

One of our primary science objectives is to probe the structure of mass loss at the times when the lobes and rings  first formed through measurements of  the proper motions of these features.  This science objective is plausible because, as we assert next, the structures in CRL\,2688 seen today accurately mimic their geometry at the time of their formation.  First, we note that pPNe do not show signs of extended nebular emission lines from species such as H$^+$ (though a few show low-ionization emission lines from localized shocks). This is a crucial observation since the onset of nebular ionization drives strong D-type ionization fronts into the neutral gas that are preceded by an expanding supersonic ``piston'' that can quickly sweep up, displace, and shock heat the upstream gas. 

Additionally, unlike mature PNe, there are no traces of  rapidly expanding central cavities that are formed by fast winds ($\sim10^3$ km s$^{-1}$) that create a hot, sparse central ``bubble'' ($10^6 - 10^8$K). The supersonic growth of the bubble displaces and compresses the gas that once occupied its interior (e.g., \citealt{1978ApJ...219L.125K}, \citealt[and subsequent papers]{2005A&A...441..573S}).  Also, there is no evidence of any other forces (e.g., strong embedded magnetic fields) or substantial shock-induced radiative energy losses that impede or alter the structures that have been observed in pPNe.  Therefore it comes as no surprise that the dense, bright cores of pPNe are found to expand uniformly (\citealt{2001A&A...373..932A, 2002A&A...386..633C, 2004ApJ...617.1142S} \& \citealt{2007A&A...468L..41A}). In other words,  the initial geometries and ejection speeds of the lobes and rings of CRL\,2688 are preserved in the reflection nebulae observed now.   

There are very few previous studies of proper motions in pPNe and none of their rings.  \citep{2001ApJ...550..778B} were the first to study proper motions in pPNe.  They used a pair of images separated by $\sim3$ y to monitor the apparent  motions of knots near the symmetry axis of Hen~3-1475.  CRL\,2688 has no counterparts of these knots.  Of more direct relevance, \citet[hereafter ``U06'']{2006ApJ...641.1113U} used a pair of HST NICMOS images of CRL\,2688 at 2$\micron$ separated by 5.5 y (unspecified camera and spatial resolution) to study proper motions of the N--S lobes.  U06  derived an inclination $i$ of $7.7^\circ$ from the plane of the sky, a distance D of 420 pc, a stellar luminosity L of 10$^{3.5}$ L$_{\odot}$, a total shell mass M of 1.2 M$_\odot$, and an expansion age T$_{\rm{kin}}$ of 350y\footnote{ Note that other inclinations and distances 16$^\circ$ and 1 kpc, respectively, have been suggested by \citet{1984ApJ...278..186Y}).}. The north lobe is tipped forward.  At this distance $1\arcsec$ corresponds to 6.3 $\times 10^{10}$ km = 0.0020 pc.

CRL\,2688 is close enough (see below), many of its structures are sufficiently bright and sharp, and its foreground extinction is small enough (A$_{\rm{V}}$ = 0.88 mag) that if a feature has an in-sky motion of 10 - 20 km s$^{-1}$ then changes in the structure of its rings can be monitored at high spatial resolution with telescopes such as the Hubble Space Telescope (``HST'') over $5 - 10$ y.  In this study we used high-resolution images from 2002 and 2009 at $0.6\micron$ from the HST archives. Compared to the NICMOS images used by U06, our data have a smaller diffraction core (FWHM $\approx 0\farcs067$ after dithering vs $\ga0\farcs2$, depending on NICMOS camera), lower readout noise and dark current (i.e. better sensitivity to the faint rings), and a slightly longer time base (6.65 y).  Indeed, we successfully measured the aggregate proper motions of sharp features in the rings as well as the lobes.  The images are described in \S2, analyzed in \S3, and combined with complementary $^{\rm{12}}$CO data in \S5.

A second focus of this paper is an analysis of the scattered light of CRL\,2688 as observed in four colors, 0.6, 0.8, 1.1, and 1.6$\micron$ in 2009.  These images are used to constrain a new and detailed model of radiative transfer by dust in order (1) to infer properties of the dust and its distribution and (2) to fit the spectral energy distribution (``SED'') of CRL\,2688.  See \S4.  There have been two prior attempts of radiative transfer modeling of CRL\,2688 \citep{1997A&A...328..290S, 2000A&A...354..657L}.  Skinner et al. used a dusty outflow with hollow polar lobes, but their geometry was not constrained by high resolution imaging that emerged shortly thereafter (S98a, \citealp[hereafter ``S98b'']{1998ApJ...493..301S}, and \citealp{2000ApJ...531..401W}).  Lopez \& Perrin proposed a model based on a dust density law smoothly decreasing with latitude from the midplane.  However, deep images of CRL\,2688 do not support this geometry.  

Our major findings are summarized in \S6 and a brief discussion of viable ring formation mechanisms is presented.

\begin{figure}[H]
\begin{centering}
\epsscale{1.0}
\plotone{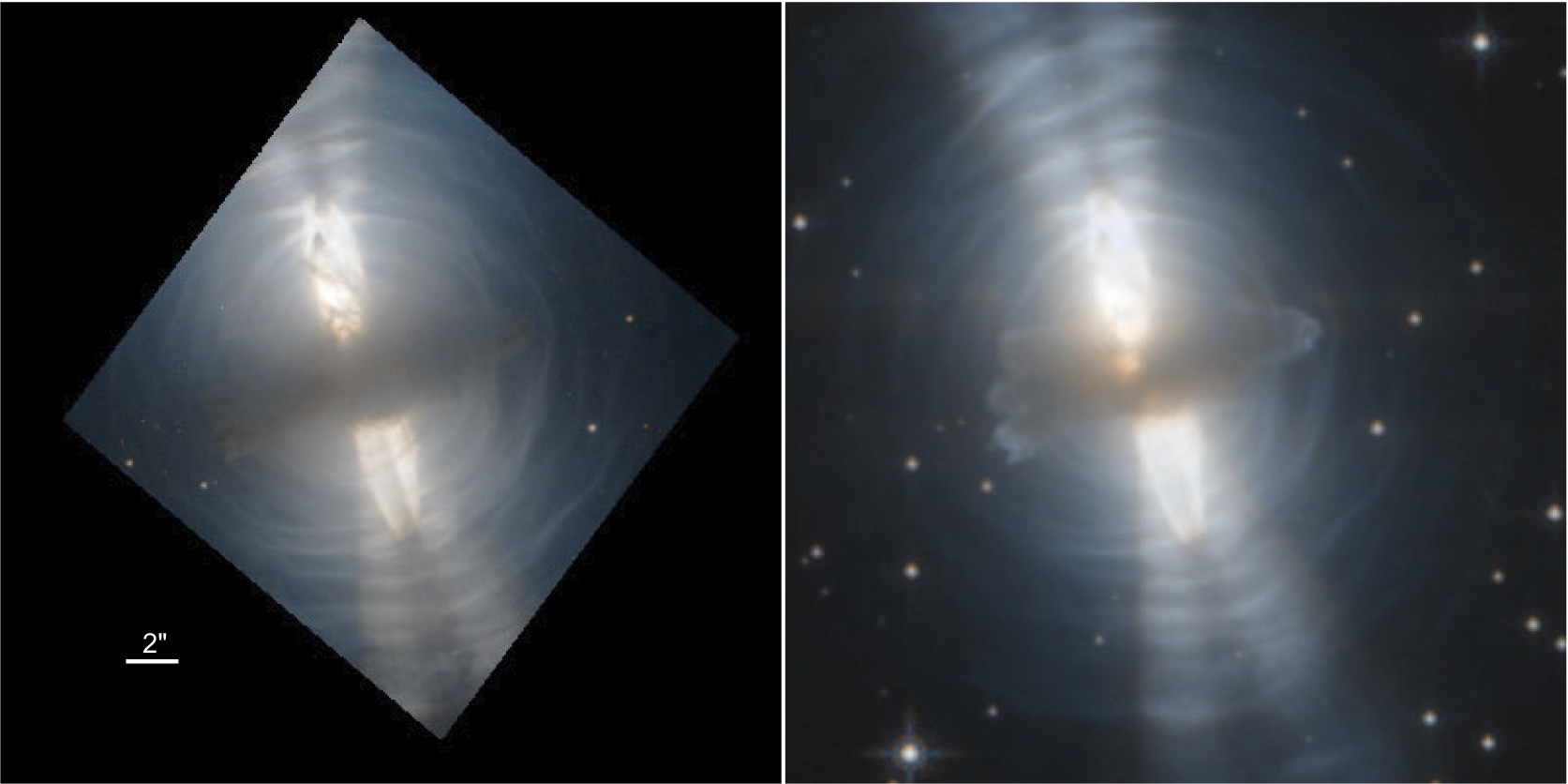}
\caption{Color overlays of the F606W-F814W (left) and F110W-F160W images (right) shown to the same scale and pixel size (0\farcs08) with a highly non-linear contrast function and exaggerated color saturation.  The native resolutions of the images are 0\farcs040 (visible) and 0\farcs128 (IR).  The FOV is 400 pixels or 32\arcsec. The color saturation has been enhanced to highlight the color changes.}
\label{Color}
\end{centering}
\end{figure}

\section{Data Calibration}

We obtained dithered images of CRL\,2688 using WFC3 (Wide Field Camera 3 Instrument Handbook, Version 2.1, Dressel et al. 2010) through four broad-band filters, F606W, F814W, F110W, and F160W (Table 1.)  
Detailed characteristics of the camera and its filters are described on the WFC3 web pages\footnote{  \url{www.stsci.edu/hst/wfc3}}.  We obtained dithered short- and long-exposure images through each filter.  The F606W and F814W exposures were limited to a single HST orbit in August 2009.  F110W and F160W exposures were obtained during another orbit in October 2009.  All exposures were made with a 512$\times$512 subarray in order to expedite data transmission to the ground.  The visible images were initially drizzled onto a $0\farcs0396$ grid, the native pixel size of the UVIS camera on WFC3.  A proper motion of 10 km s$^{-1}$ per decade corresponds to the width of a full F606W-F814W pixel at the assumed distance of CRL\,2688 of 420pc.   The infrared images were drizzled to a grid of $0\farcs128$.  A few saturated pixels in some images of the lobes were replaced with unsaturated values obtained from short exposures.  

Color overlay: The color distribution of scattered light was measured from flux-calibrated WFC3 images in each filter after using bright field stars to align each image.  The flux calibration is based solely on calibration data supplied with the images from the standard calibration pipeline. The basic results are shown in Fig \ref{Color}.  In this figure the color saturation and contrast were altered in order to exaggerate the brightness of the faint outer rings and to emphasize the overall color variations. 

Proper-motion studies: Only images in the F606W filter were used for measuring proper motions since the archival data in this filter are the highest in spatial resolution and signal-to-noise ratios. Table 2 
shows a log of the F606W images including exposures dates and duration and the original pixel sizes and fields of view. Detecting changes in structure based on images originally made using different cameras at various epochs requires thoughtful image alignment and gridding, the best possible corrections for camera sensitivity variations and geometric image distortions, careful image registration, and well-matched point-spread functions. We describe our procedures next.

In every case we used the latest corrections supplied by STScI staff in May 2010.  Note that geometric distortion in the central $512^2$ subfields of the UVIS and IR cameras are very small.   Distortions in the ACS/HRC camera have been carefully measured since 2002.  Corrections for distortion were applied to the pipeline data from the archives using Multidrizzle (described in the January 2009 version of the handbook\footnote{ \url{www.stsci.edu/hst/HST \_overview/documents/multidrizzle/multidrizzle.pdf}}.  Based on HST documentation we believe that the systematic errors in the corrections are generally much less than 0.1 pixel (we have no means of confirming this result). In short, large-scale spatial distortions that might materially affect our results should be insignificant.

Image alignment is tricky since no suitable stationary point-like sources are found in the field of the CRL\,2688 images.   That is, each of the field stars in the images have idiosyncratic motions relative to CRL\,2688. As noted earlier, the nucleus of CRL\,2688 is not visible within any of our filter passbands.  Thus the images from each epoch and filter were initially aligned from the absolute positions of the images that are stored in the image headers. (These coordinates are based on guide stars, each with its own proper motion.)  Then image pairs were aligned empirically using the centers of symmetries of large-scale features, such as the lobes and arcs. We are confident that the uncertainty of the image alignment,  $\approx0\farcs005$, is a small fraction of the PSF of the cameras (FWHM = 0\farcs067 after dithering).  The nebular symmetry center agrees closely with the polarization center \citep{2000ApJ...531..401W} and the $^{12}$CO kinematic center (C00).

We searched for proper motions of the structures within CRL\,2688 in all image pairs.   All of our pair-by-pair results and our assessment of their uncertainties  are compiled in Table 3.  Note that CRL\,2688 was observed using the low-resolution  ($0\farcs1$/pixel) WF3 chip in WFPC2 in which the PSF varies noticeably with chip location.  Unfortunately, the 1995 image is not sufficiently deep for reliable results of changes in the positions of the faint outer rings. Thus the result for the 1995-2009 image pairs in Table 3 is based largely on the innermost rings.  What's more the camera focus appeared to be poor since we had to convolve the 2009 images to a  PSF of 0\farcs2 to match the PSFs of field stars in the 1995 image.  Using the 1998 and 2009 image pairs produced rather somewhat irregular inconsistent patterns of nebular ring growth.  In short, our assessment of the systematic errors led us to concerns about the quality of the results obtained using any of the early WFPC2 images for studies of large-scale proper motions that our error estimates may not capture.

The results derived from ACS/HRC (c. 2002.94) and WFC3 images (c. 2009.59) are the most credible and the only ones discussed hereafter (6.65y time base).  Those images were first drizzled onto a $0\farcs025$ grid (the native pixel size of the ACS/HRC image) and the intensities were normalized to the fluxes of three bright field stars. As a check, we confirmed that this image renormalization resulted in an image ratio that is very close to unity in the smooth, bright portions of the lobes, as expected.  Therefore any residuals from multi-epoch image subtractions are not artifacts of errors in camera flux calibrations.  

Finally, the 2002 and 2009 images were converted from counts per second to units of energy flux, divided (after suitable magnifications to correct for nebular expansion), and examined for changes in surface brightness.  The conversion to energy flux is based on the value of PHOTFLAM stored in the image header.  PHOTFLAM accounts for the system and filter throughput and the widths of the filters.  Values of PHOTFLAM are based on methodical calibrations of the system by STScI staff and are very stable for the F606W filters used in our study.

\section{Analysis \& Results}

\subsection{Image structures and color variations}

The structure and color variations of CRL\,2688 were presented in Fig \ref{Color}.  The most prominent new features are these:
\begin{itemize}

\item the southern lobe and the pairs of ``searchlights'' that continue beyond its tips are noticeably redder than the northern lobe and its corresponding pair of searchlights.  
\item the opening angle of the searchlights widens and the gaps in their centers close noticeably at longer wavelengths.  We surmise that the gap is formed by a shadowing obstacle whose edges become more transparent with increasing wavelength. 
\item the rings are generally bluer in color than the lobes. One exception is where the rings cross the southern lobe that is tipped backwards.   
\item the dark dust lane and especially the finger tips at its extremities are outlined in continuum light in our $1.1\micron$ image.  This outline is far less prominent in our $0.8\micron$ and $1.6\micron$ images.
\item the north (south) lobe faded by 10-12\% (2-4\%) between 2002 and 2009.  The E--W dust faded with an irregular pattern. The northern (southern) quadrant of the lobes faded by 15-20\% (5-15\%).  See Fig 2b and 2c for the pattern of brightness changes between 2002 and 2009.
\end{itemize}

N--S lobes: The lobes are flame-shaped emitters that both absorb and scatter starlight radiated into their base.  As such, they are probably the main source of illumination for the surrounding ensemble of rings at lower latitudes.  

The north lobe is marbled with deep extinction filaments (most with H$_2$ emission counterparts).  Its relatively blue colors are partially attributable to the shorter interior line of sight to its forward-tipped lobe.  Also, as pointed out by \citet{1980PASP...92..736C},  the north (south) lobe is seen in forward- (back-)scattered light.  It will appear bluer if the scattering phase function is not isotropic. 

The fainter southern lobe is generally smoother in appearance than its northern counterpart and devoid of marbled dust.  It divides cleanly into two mirror-symmetric features separated by a gap along its symmetry axis that extends into the searchlights.  We surmise that the south lobe is hollow, or that the interior gap along its symmetry axis is an illumination shadow cast from the E--W dust lane, or both.  The continuous dark gap extending from the base of the southern lobe into the searchlights support the shadow hypothesis without eliminating the idea of a hollow lobe core.  It is difficult to tell whether the north lobe is similar in this regard.

Searchlights: The bright searchlights are best explained if the lobes are semi-transparent to starlight when viewed from very high latitude.  The red gaps in the searchlights are shadows, as proposed by S98b and others. Each searchlight has a dark core along the nebular symmetry axis.  As proposed by S98b and \citet{2007apn4.confE..55L}, H$_2$ knots at the tips of each lobe may be occulting disks that cast shadows into each searchlight.  In practice, the exact location of the symmetric pair of dark occulting disks cannot be located in any of our images.  They could be anywhere on the symmetry axis, perhaps even somewhere inside the dark E--W dust lane.  

E--W dust lane:  We noted that the edges of the dust lane are outlined nicely in our $1.1\micron-$ and, to a lesser extent, in the 0.8 and $1.6-\micron$  images.  The scattered light follows essentially the same brightness pattern as the shock-excited H$_2$. As noted above, the fingertips are especially bright.  Given the faintness of the rings at low latitude, it seems unlikely that the fingertips are illuminated from the exterior.

Close to the geometric center of nebular symmetry the present infrared images show two compact and very reddened features  that have no visible counterparts.  Both of these lie near the outer edge of the central CO cavity of $2\arcsec$ in diameter.  The brighter of these IR features was identified as ``knot A'' by G02 in their studies at 3.8 and 4.7$\micron$.  The same knot appears in  Fig \ref{Color} and was seen at 1.65 and 2.15$\micron$ by S98a.  The fainter and redder of the two near-nuclear infrared features aligns with a feature at the base of the southern lobe reported by G02.  These features are also very proximate to the presumed locations of the two ``polar caps''  in the scattering model to be described in \S4.

\subsection{Growth Patterns of the Lobes}

Methodology: Monitoring changes in the locations of sharp structures must be tailored to the nebular geometry; e.g., changes in a circular feature involve finding a change in its radius.  Our efforts to fit the lobe expansion rely on the sharpness of the edges of the lobes and the filamentary dust features in their interior. The earlier image of a registered pair was magnified a factor $(1 + M)$, where $M \ga 0$ is a free parameter.  The ratio of the magnified earlier image to the newer image of the pair produces a ``ratio image'' in which radial growth is seen as large-scale dark-light features wherever sharp gradients  in intensity are located.\footnote{ As a rule, we found that the image ratios are a better measure of alignment than are the differences, at least for the outermost rings.  This is because the surface brightness falls off rapidly with radius,. So the pattern of differences at the outer edges of the images become very faint.  However, the rings can be traced to the noisy outer edges of a ratio image.}

The signal of structure change is clear when no magnification is applied ($M = 0$).  $M$ was incremented until the systematic signals of change were minimized (i.e., ``flattened'').  The goodness of fit is measured by eye since standard statistical measures of variations are dominated by readout noise, hot pixels, and the strong residuals left by translated field stars.  We found that for the eye, a small range of values of $M$ flattens the residuals equally well.  This range determines the uncertainty of the fit.

\begin{figure}[H]
\centering
\epsscale{1.0}
\plotone{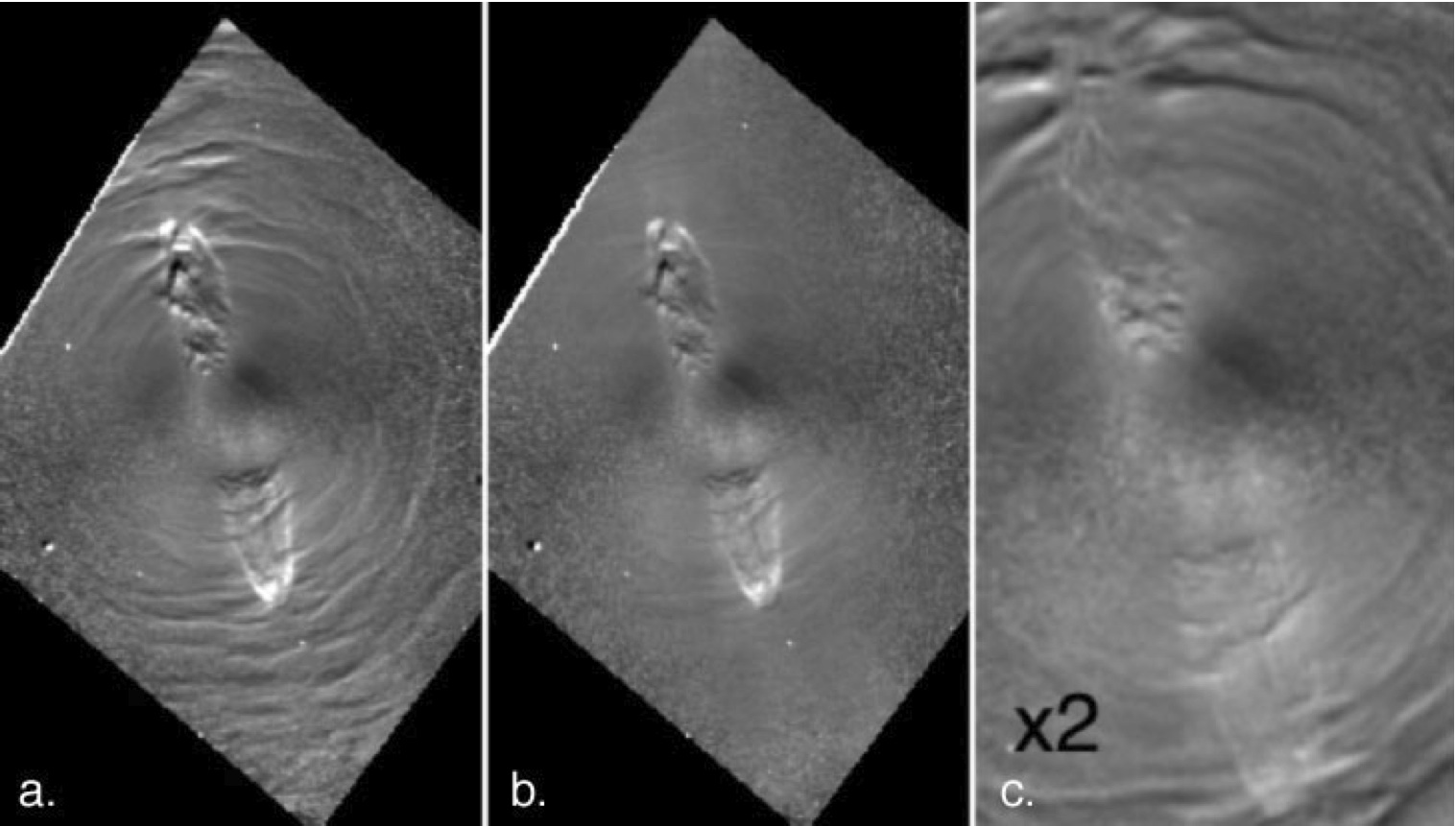}
\caption{Ratios of flux-calibrated HST images of CRL\,2688 in the F606W filter in 2002.94 (ACS/HRC) and 2009.59 (WFC3). The intensity display scale ranges from 0.5 (black) to 2.0 (white) in all images.  North is up. Left panel ($M = 0$): the ``signal'' of structural changes .  Middle panel ($M = 0.009$): best overall flattening of the rings using the magnification method.  Right panel ($M = 0.025$ and the image scale has been doubled): best overall flattening of the lobes using the magnification method.  Note that the expanding rings are absent (prominent) in projection on the north (south) lobe implying that the lobes are largely opaque at $0.6\micron$.}
\label{ratio}
\end{figure}

Results: The residual images are shown in Fig. \ref{ratio} for three values of $M$ applied to the ACS/HRC image of 2002.94 prior to forming the ratio image: (a) no magnification $(M = 0)$, (b) $(M = 0.009)$ and (c) $(M = 0.025)$.  In the case of the lobes the residuals in the ratio image are very nicely flattened for $M = 0.025 \pm 15\%$.  In other words,  $\Delta\theta_r \propto \theta_r$, where $\Delta\theta_r$ is the radial offset of a lobe feature at a radial offset form the symmetry center at $\theta_r$ = 0.  (N.B.:  We were unable to tell whether the same value of $M$ applies to the expansion of the edges of the E--W dust lane.)

This analysis yields an expansion age for the lobes of 6.65y/0.025, or 266y $\pm 15\%$ in good agreement with the estimate of (1) 350y found by U06 based on their comparison of lower-resolution HST/NICMOS images spanning 5.5y, (2) 200y found by \citet{1990ApJ...351..222J} for the time since the star left the AGB (assuming a distance of 1 kpc), and (3) the CO kinematic ages of 52 to 500 y for various pairs of CO knots embedded in the lobes (C00 after adjusting to D = 420 pc).  Earlier age estimates require a prior knowledge of the inclination angle $i$ and a distance D and, hence, are less accurate.\footnote{ Ages derived from proper motions are not affected by uncertainties in $i$ or D.  However, age estimates based on measured Doppler shifts are very sensitive to any errors in D and $\sin{i}$, especially when $i$ is small. U06 estimated a lobe inclination of $i = 7.7^{\circ}$ but could not rule out larger values. All prior estimates of $i$ and D vary from those of U06 up to a factor of two \citep{1984ApJ...278..186Y}.  We note that $i \approx 15^{\circ}$ is nicely consistent with our scattering model in \S4.}

\begin{figure}[H]
\epsscale{1.0}
\plotone{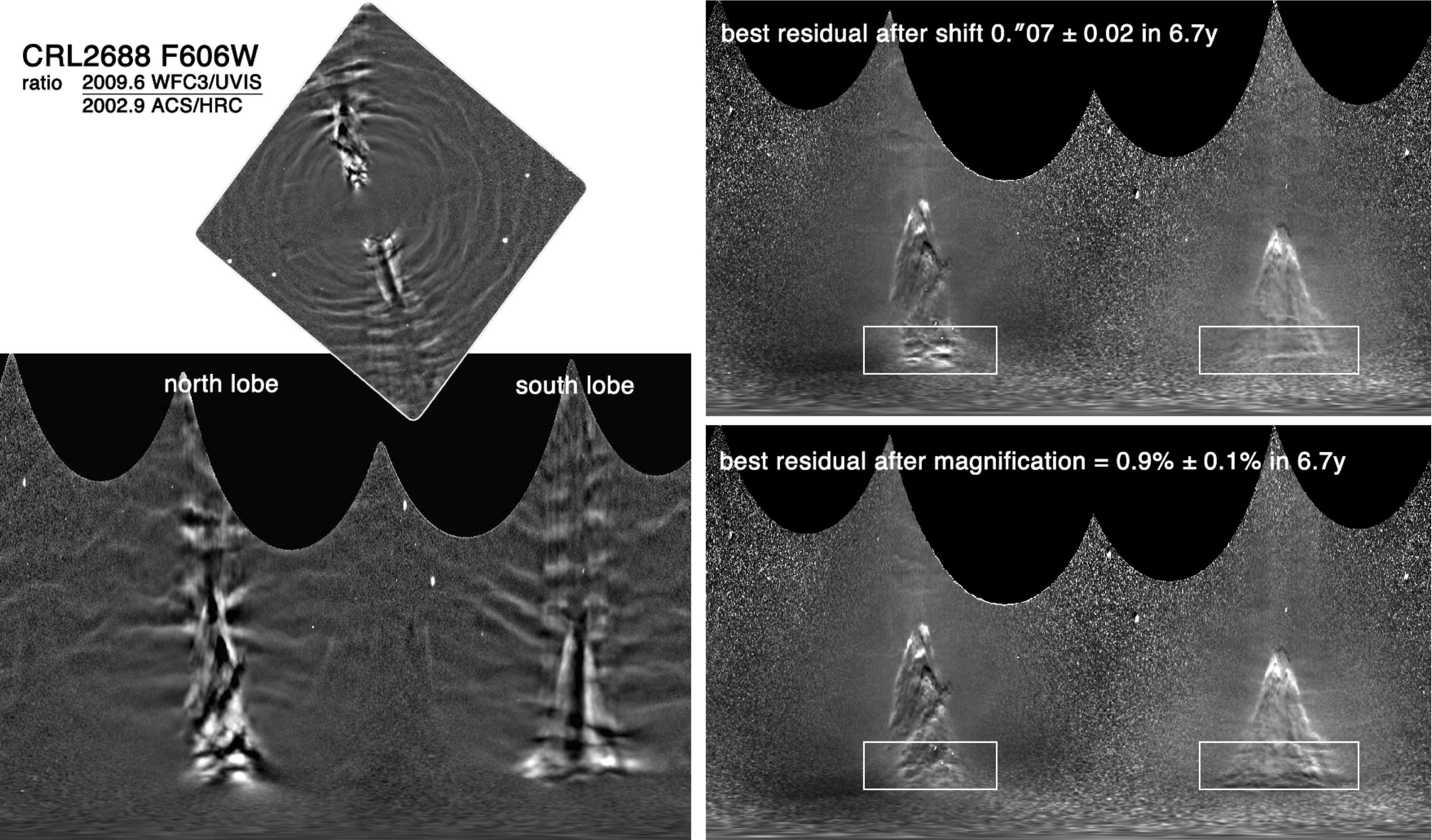}
\caption{Ratio images of CRL 2688 shown in polar coordinates.  Left: unsharp-masked ratio image in Cartesian (upper) and polar coordinates (lower), and assuming $M = 0$.  Top right: as lower left except for a radial translation of the 2002.94 image by 0\farcs07 prior to forming the ratio image.  Lower right: as lower left after a magnification of the 2002.94 image by a factor of 1.009.  Although the rings are well flattened throughout most of the lobes in the upper and lower images, faint ``ghosts'' of the rings are more prominent within the white box.}
\label{polar}
\end{figure}

\subsection{Growth Patterns of the Rings}

Methodology:  We investigated two models for fitting the growth pattern of the rings: (1) a brief epoch of ring  ejection followed by  a homologous ``uniform expansion'' in which $\dot\theta_r \propto \theta_r^1$ and (2) ``ongoing ring ejections'' with constant motions in which $\dot\theta_r \propto \theta_r^0$ for each ring.  The former suggested itself during image alignment, as noted in \S2.  Uniform expansion is the most common Doppler expansion pattern in many other pPNe and PNe (\S1).  On the other hand, the ongoing ring ejection is a popular but thus-far untested paradigm in which an AGB star is somehow regularly stimulated to eject a bubble of enhanced density or a new compression wave every $10^2-10^3$y (\S1) at the same speed (so that the rings do not overtake one another and lose their separate identities). As noted in \S1, this idea is motivated by the ensembles of concentric rings seen in many other pPNe and PNe.  

In order to find the expansion pattern of the rings, the observed images were transformed into polar coordinates and their alignments carefully checked before forming the image ratio.  As seen in Fig. \ref{polar}, circular rings and arcs become horizontal lines. On the other hand, if the rings each have the same ejection speed $v_r = \rm{D}\dot\theta_r$ then the entire system of rings will translate by a fixed amount $\Delta\theta_r$ in the radial direction between observing epochs.  Thus the residual will be best flattened by applying an optimum choice $\Delta\theta_r$ for the radial translation of the ring ensemble of the earlier image prior to forming the image ratio.

Results:  Both models of uniform expansion fit the WFC3 data well.  We find that the ratio image is optimally flattened everywhere if we adopt $M = 0.009 \pm 0.002$.  This value of $M$ corresponds a the kinematic age of the ensemble of rings is $\Delta t_{kin} = $ 6.65y/$M$ = 740y with 20\% uncertainty.  A very good fit is also found  for the constant ring ejection-speed model. A translation $\Delta\theta_r = 0\farcs07 \pm 0\farcs01$ nulls most of the ensemble as seen in Fig. \ref{polar}.  However, strong residual structure is quite conspicuous on both sides of the dark E--W dust lane within the white box shown in this figure.  (Ignoring this region presumes that the structure within the white box is part of a different kinematic system such as a lobe or the adjacent dark dust lane.)
$^{12}$CO observations discussed in \S5 will break this degeneracy in the quality of the models.

As noted in \S1, a study of the growth pattern of the rings is a key result of this paper.    However, the present study is limited to arcs within $\approx20\arcsec$ of the nucleus.  Therefore an analysis of multi-epoch images spanning the full $\approx80\arcsec$ field of the rings seen in the  ACS/WFC F606W image of 2002 would be a very useful next step for a better description of the ensemble expansion pattern and the ejection histories of the rings of CRL\,2688.

\section{Model for Scattered Light}

\subsection{Overview and Previous Results}

The most conspicuous features of the reflection nebula CRL\,2688 seen in HST images are its highly articulated and symmetric shape.  The WFC3 images trace only scattered starlight, some of it multiply scattered.  Clearly this provides a highly biased view of the actual density distribution.  Therefore a better assessment of the spatial distribution of dust requires a realistic analysis in which starlight computationally transferred through a series of dusty features of different geometries.   In this section we describe a successful model in which each structural component is generally characterized by the geometric features in our images and the dust is described by particles of various sizes, size distributions, and scattering and absorption properties.  Details of the model can be found in Appendix A.

\subsection {Model Methodology, Strategy, and Assumptions}

Constructing a model that simultaneously reproduces the complex structure, color variations, and SED of CRL\,2688 is challenging.   We have employed the radiation transfer code LELUYA \citep[see also \url{www.leluya.org}]{2003PhDT........40V}. LELUYA is a powerful yet flexible tool which nicely solves the integral equation of the formal solution of radiative transfer including dust scattering, absorption and thermal emission in two spatial dimensions. The transfer equations are solved on a highly unstructured triangular self-adaptive grid that  simultaneously traces both the density gradients and the optical depth gradients over many orders of magnitude in spatial and optical depth space. It allows the user to assign grains of various sizes and particle distributions to each feature.  However, LELUYA does not treat polarization and non-isotropic scattering.

We use spherical amorphous carbon dust grains \citep{1988ioch.rept.....H}. The optical constants for these grains are derived by presuming isotropic scattering and Mie theory.  For simplicity we  assume that the chemical properties of the dust particles are uniform; i.e., only their size distribution varies from feature to feature.

\begin{figure}[H]
\epsscale{1.0}
\plotone{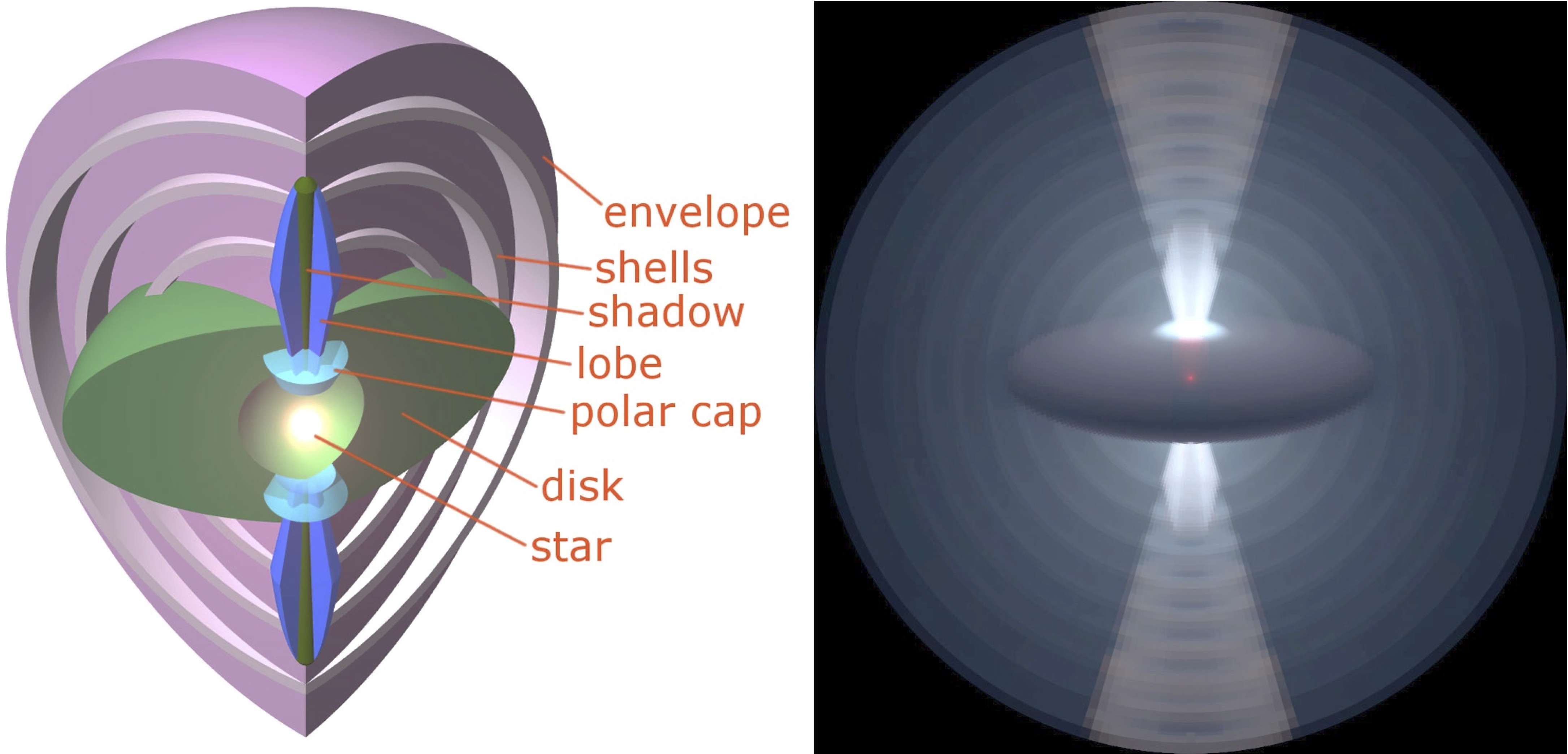}
\caption{Left: sketch of the 2D model for the circumstellar dusty environment of CRL\,2688. All components except the "polar caps" are constrained in their shape by imaging data. Each of the model components is described in Appendix A. Components in the sketch are not shown exactly to scale; for the list of their parameters see Table 4.  Right: color rendition of the model images with a non-linear contrast, blue $= 0.6\micron$, green $= 1.1\micron$, red = $1.6\micron$.  The emerging light from the nebular core at $4.5\micron$ is overplotted in red.}
\label{DejanA}
\end{figure}

Constrained and unconstrained model variables:  Fig. \ref{DejanA} provides a conceptual sketch of the structural components used for the computations. We tried to specify as many modeling constraints from the observed four-color images (\S2) as possible in order to reduce degeneracy of the results of the model.  However, the core of the nebula where the illumination originates is hidden from direct view even beyond 4$\micron$ (G02) so its structure was initially assumed and then optimized to fit the SED and the patterns of scattered light.  We next describe how the many model variables were fixed or fitted.
  
The dust distribution is shown schematically on the left side of Fig \ref{DejanA} and described in detail in the Appendix. A total opening angle of $30^\circ$ was adopted for the lobes and the searchlights, $10^\circ$ for the shadow zone along its symmetry axis, and $40^\circ$ for the polar caps.  The density profile adopted for the caps, the ellipsoidal equatorial disk, the hollow lobes, and the envelope scales as $r^{0}$, $r^{-0.5}$, $r^{+1}$, and $r^{-3}$, respectively.  

The geometry and density of the polar caps and the radiating area of the star were varied in order to match the SED and color distributions of the structures.  We adopted two models for the illumination source with different temperatures, 4000\,K and 7000\,K.  Scattering and absorption of starlight was then followed through the dust features. The particle-size-dependent opacities are taken from \cite{1988ioch.rept.....H}.  

Other features are those identified in S98a and the circumstellar cavity noted in C00.  The  polar caps enclose this cavity close to the star where dust particles initially form.  The stellar illumination is launched through annular openings in the caps on its way to the lobes and the searchlights.  A distance 420 pc was assumed (\S1).  We treated $i$ as a free parameter to be determined by assuming that the two lobes are of equal brightness and given the dust distribution in the model envelope, tilting the lobes to match their observed brightness ratio.  

It is worth noting that we are unable to derive even an approximate dust mass for several reasons. One is that the grain sizes (specifically, the scattering area per unit dust mass) are not tightly constrained by the model.  Also, we assume a uniform grain radius or size distribution for each structural component.  This is obviously unrealistic.  Moreover, we do not try to incorporate scattering by large dust particles into the model.  \citep{2000ApJ...528L.105J} estimated that the large grains could be at least as large as 5 mm.  Such large grains may have a major impact on the derived dust mass.

\begin{figure}[H]
\epsscale{1.0}
\plotone{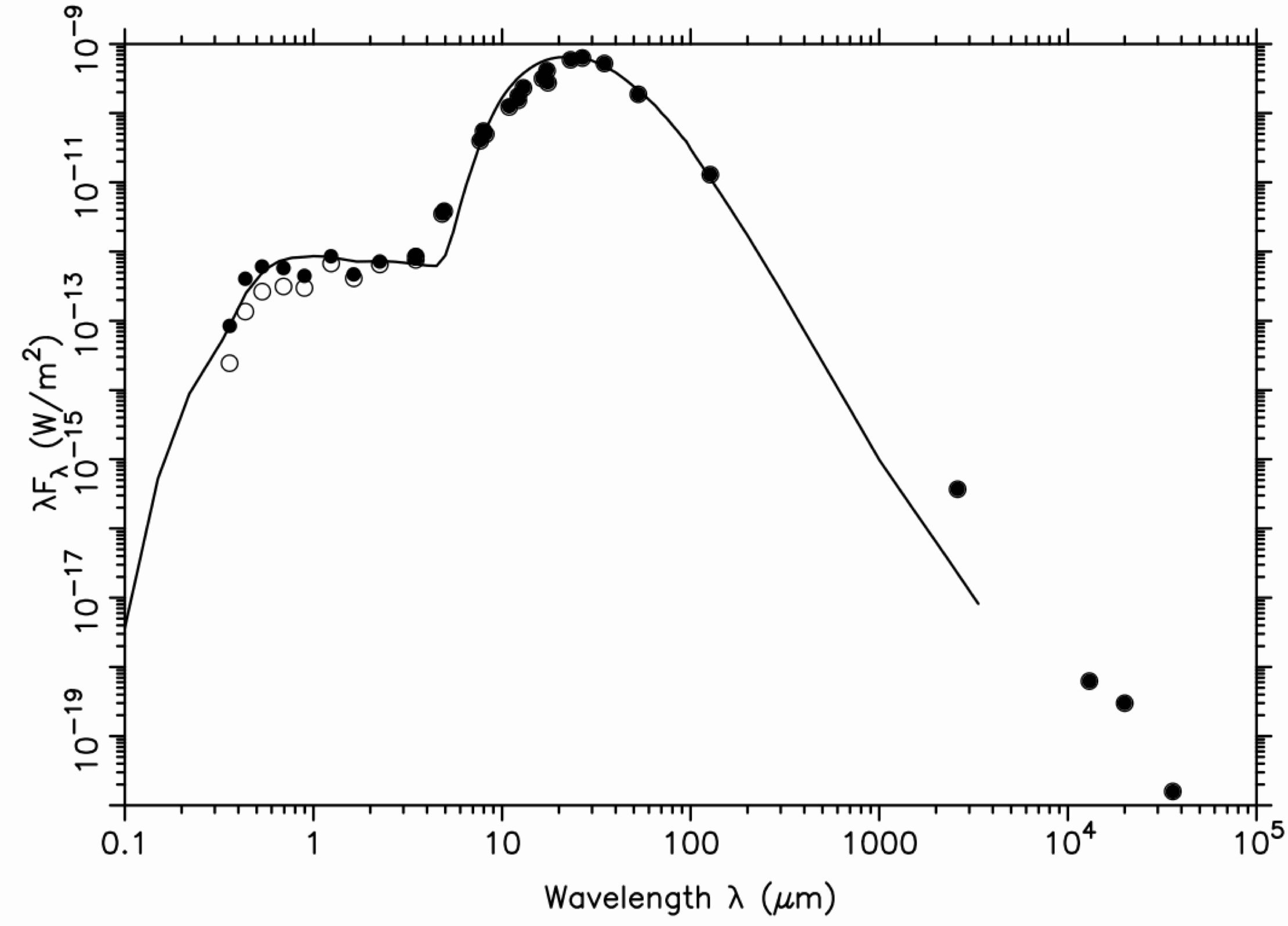}
\caption{A fit of the predicted SED with observed data points before (open circles) and after (solid circles) corrections for foreground extinction of A$_{\rm{V}} = 0.88$ (U06).}
\label{DejanC}
\end{figure}

\begin{figure}[H]
\epsscale{1.0}
\plotone{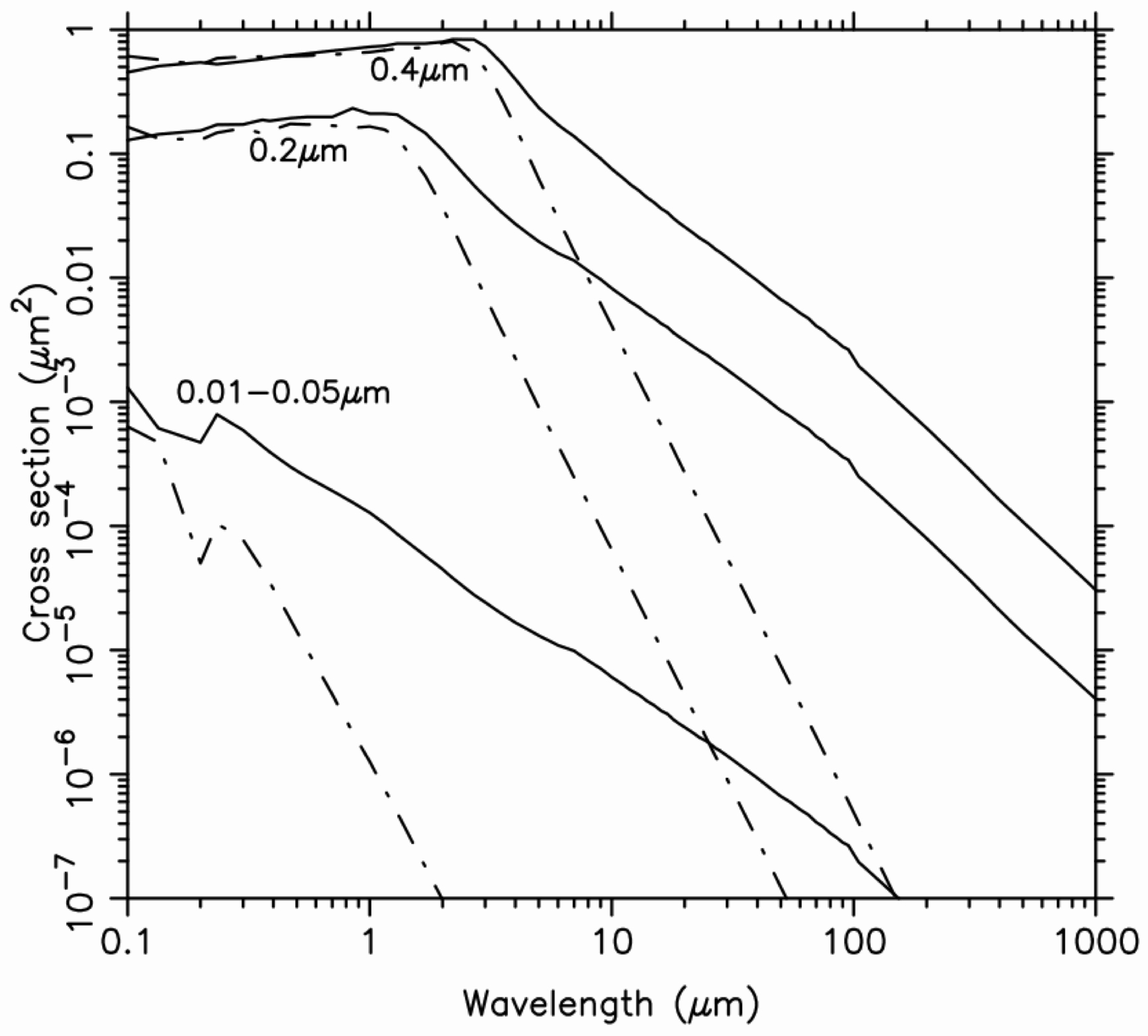}
\caption{Cross section of dust grains used in our model. Solid lines are absorption, dash-dot lines are scattering cross sections. Grain radii are indicated in the figure. A distribution of grain sizes for the smallest grains (lower curves) is a power law of exponent -3.5.}
\label{DejanB}
\end{figure}

\subsection{Model Results}

The success of the model in fitting the data is shown in Figs. \ref{DejanA} (right side) and \ref{DejanC}.   Generally speaking, the overall correspondence of the geometry and color variations of the model and our HST images is excellent.  Of course, much of this is the direct result of the flexibility in designing the density and dust particle size distribution of the model.  The resulting color variations are in good agreement except (arguably) in the faintest outer rings and searchlights where the quality of the images is poorest.   The modeled arcs show the basic morphological properties of imaged arcs, even though we use ideal spherical shells in our model.  We confirm the estimates by \citet{1998ApJ...493..301S} that arcs represent a local density increase by a factor of 3. A key feature of the model is that the SED is reproduced very successfully across the spectrum from the visible to the sub-mm. 

We note that the long run time of the model made a detailed parameter study of many parameters impossible. Thus it is difficult to assess the uniqueness and the uncertainties on the selection of all of the parameters in the model. 

N--S lobes and E--W dust lane: We find that the optical opacity $\tau$ of the lobes $\approx2$ at $0.55\micron$ and that the opacity of the disk is $\approx$55 in its midplane. The model image is shown at an inclination angle $i = 15^\circ$ . This value of $i$ best compensates for the unequal observed surface brightness ratio of the north and south lobes (assuming that both lobes are identical and that scattering is isotropic).   The result agrees with the value of $i = 16^\circ$ derived from the scattering model for CRL\,2688 by \cite{1984ApJ...278..186Y}.  Even so, the fit to the image morphology and SED is adequate for $5^\circ \la i \la 15^\circ$. A list of other derived model parameters is provided in able \ref{table-fit} of Appendix A.

Central star: We obtained a bolometric flux of $1.0 \pm 0.2 \times 10^{-9}$ W m$^{-2}$.  This result is based on a correction for asymmetric diffuse radiation. The correction arises from the lack of spherical symmetry which is typically ignored in the literature when the luminosity is calculated. That is, more energy is visible from a vantage point along the symmetry axis (large inclination angles) than near the midplane where the disk is blocking the central star (small inclination angles). Our model takes this into consideration. A detailed discussion of this effect is given by \cite{2004MNRAS.352..852V} in their \S{A1}.

We estimate the light originates from a star with a luminosity of $5500 \pm 1100$ L$_\sun$ and a temperature T$_* \sim$ 7000 K, in close agreement with \citet{2000AstL...26..439K} assuming D = 420 pc and A$_{\rm{V}} = 0.88$. The corresponding stellar radius is $0.23 \pm 0.02$ A.U. ($51 \pm 5 R_{\sun}$), or $0.56$ msec $\pm 10\%$.

Circumnuclear core:  The radiative transfer calculation yields the radius of the dust-free cavity of 46.4 stellar radii, deep within the CO cavity.  This value critically depends on the dust condensation temperature, which is in our case T$_{\rm{max}}$=1000K in the midplane.   Since this value is not a measurement, and because models with other temperatures are viable, we adopted this value to reduce model degeneracy.  Newly formed dust radiates in the near IR spectrum which is completely obscured by the dust lane.  Thus photometric verification of the size, temperature, and radiated flux of the is not amendable to empirical verification.

To first order, the computed dust temperature distribution is dominated by a simple gradient outside of the polar caps near the star.  The temperatures in all directions drop smoothly from 1000\,K at $\log(r/R_{in})\sim0$ to $300$K at $\log(r/R_{in})\sim1$ and $100$K at $\log(r/R_{in})\sim2$. Here $R_{in}$ is the dust condensation radius (the radius of the inner grid).   The dust temperature continues to fall smoothly to the edge of the entire grid at $R_{out}=600R_{in}$. 

With this cavity size and assumed limit on the condensation temperature we can rule out the maximum dust temperature of 330K as suggested by \cite{2000A&A...354..657L}. Asymmetric diffuse radiation and differences in grain sizes create variations in the highest dust temperature on the spherical surface of dust-free cavity. The large $0.4\micron$ grains in disk and lobes experience temperatures from 1000K at the midplane to $\sim$1100K on the polar axis.  In contrast, the small grains (0.01 -- 0.05$\micron$) in the polar caps reach 1400K since smaller grains are relatively inefficient emitters of infrared radiation.

Polar Caps: Recall that knot A seen in the infrared (G02 and Fig \ref{Color}) lies within 1$\arcsec$ of the expected location of the north polar cap.  It corresponds with the bright zone at the upper edge of the polar cap. Similarly, the fainter infrared knot lies near the expected location of the heavily obscured polar cap at the base of the southern lobe.

Rings:  Although the radiative transfer model was designed to reproduce the geometry of the rings, searchlights, and the shadow zone on the nebular symmetry axis, it does not fully capture the color variations outside of the nebular core. The HST images indicate that the searchlights are relatively blue along their full extent whereas the model shows the blue colors are found only where the nearby blue lobes dominate the illumination of the rings.   Some of this can be attributed to the two-dimensional nature of the model which cannot accurately treat multiple scatterings.  

Two other areas of model-image disagreement are (1) the bright parts of the searchlights and their relatively red edges and (2) the opening angle of the shadow within the searchlight that decreases as the wavelength increases.  These problems would probably be resolved if we model the annular aperture that defines the outer edges of both the lobes and the shadow as having ``soft'' edges with opacity gradients.

Another area of disagreement is that the disk used in the model is darker and thicker than its counterpart seen in HST images. This is readily accommodated by a change in the assumed geometry and density of the disk.  In reality, the dark disk seems to be highly structured (C00, G02 and \S3), so full agreement with a highly idealized model agreement is not anticipated.

SED: The agreement of the predicted and model SED is one of the major successes of this model.  The best fit to the SED is shown in Fig. \ref{DejanC}.  The fit starts to fall below the data at mm wavelengths where thermal radiation from large dust particles dominates.  Nonetheless, the overall quality of the fit is far superior to those of previous models \citep{1997A&A...328..290S, 2000A&A...354..657L}.

Dust particles: The dust particle size and size distribution were treated as free parameters and adjusted to provide a satisfactory match to the observed image data and the SED. The data do not tightly constrain these particle sizes.  We adopted a uniform grain size of $0.4\micron$ for the disk and lobes, $0.2\micron$ in the envelope including the rings, and a particle size distribution that scales as (grain radius)$^{3.5}$ in the polar caps.  Available imaging and photometry at $\lambda \ge 2\micron$ provides a range of constrains on dust properties, optical depths and guidance on the limits of the interpretation of our modeling results. The data for the SED were derived from observed fluxes compiled from \cite{1975ApJ...198L.129N} and \cite{1997A&A...328..290S} and then dereddened using a fixed foreground extinction of A$_{\rm{V}} = 0.88$ (G06).

\cite{2000A&A...354..657L} proposed a different radiation-transfer model to explain the features of CRL\,2688 based on a dust density law smoothly decreasing with latitude above (and below) the midplane. Their modeled images showed brightness spikes in place of searchlight beams, which is an interesting result considering the simple configuration of their dust geometry. However, these spikes and the overall nebular geometry are simply not in agreement with the searchlights seen in HST images.  Even so, one of their firm conclusions is that the highest dust temperature is about 330K, far below the dust condensation temperature and inconsistent with the nebular SED.

\section{Comparison to CO Distribution and Kinematics}

In this section we combine our results with radio CO studies in order to develop additional information on the history of mass ejection in CRL\,2688.  As it turns out, many of the CO, the H$_2$, and the scattered-light features of the lobes and dust lane nebulae agree in location nicely, at least in outline.  Thus we can expect that outside of the circumnuclear CO cavity (see below) the volumes that generate the CO lines and the dust that scatters starlight are one and the same.  It should be noted that the present HST images cannot be used to probe the structure deep within the lobes and the dust lane since the optical opacity exceeds unity (\S4).

\subsection {Structure and Motions in the Lobes and Dust Lane}

We first consider the lobes and the dark dust lane.  In \S1 we noted that C00 investigated both the structure and the kinematics of $^{12}$CO  within the lobes and the dust lane of CRL\,2688 using a radio interferometer with $1\arcsec$ spatial resolution. They found seven pairs of finger-shaped outflows bounded by the edges of the lobes and the dust lane.  As noted in \S1, B01 decomposed the Doppler-broadened, nebula-integrated $^{12}$CO and $^{13}$CO line profiles into an unsaturated ``fast wind''  of width $\sim \pm40$ km s$^{-1}$ arising in the inner core where the finger-shaped CO outflows are found plus a saturated "slow wind" of width $\sim \pm15$ km s$^{-1}$ from the surrounding AGB wind where the rings are located.  They and C00 also found a deep and narrow $^{12}$CO absorption feature whose velocity is blueshifted by $-18$ km s$^{-1}$ from the systemic velocity of CRK\,2688. 

We summarize the high-resolution observations of the lobes and the dark dust lane by C00.  Seven pairs of fingers are revealed, each pair of which consists of outflows with nearly equal and opposite Doppler shifts, as noted in \S1. As shown in their Fig. 2, finger pairs D1-D2 and F1-F2 lie along the leading edges on the N--S lobe. E1-E2 lie at the intersection of the N--S lobes and the E--W dust lane.  A1-A2, B1-B2, C1-C2,  and G1-G2 are embedded within the dust lane.  C00 suspected that each finger may increase in velocity along its length.  Bright knots of H$_2$ and scattered starlight at 1.1$\micron$ are also coincident at the extremities of all of these fingers.  

Interestingly all eight of the finger tips associated with the dust lane also have scattered-light counterparts at 0.8 and 1.1$\micron$.    Some of the edges of the cone are illuminated by scattered light from the lobes and the rings in our HST images.  However, the outermost edges of the of the equatorial dust lane are especially prominent.  They are not likely to be illumined from the exterior since the rings are particularly faint at low latitudes.  So it's possible that starlight propagates through hollow CO fingers where it scatters from dusty clumps at the fingertips. Speculatively, the fingers (possibly outflowing jets or bullets) lie within a cone whose opening angle is $\sim 30^\circ$ -- strikingly similar to the morphology of the ensemble of finger pairs in CRL618 analyzed by \citet{2002ApJ...578..269S}.  In other words, the equatorial dust ellipsoid adopted for the two-dimensional scattering model in \S4 may instead be a pair of cones with a bundle of three pairs of CO fingers inside.

C00 derived the kinematic ages of each finger pair from the distance traversed from the nucleus and the Doppler velocity: T$_{kin}$ =  d$_{finger}$$\div$v$_{finger}$ =
(D $\theta_r$$\sin{i}$)$\div$(v$_{Dop}$/$\cos{i})$, where d$_{finger}$ is the distance traversed by each fingertip,  v$_{finger}$ is the corresponding speed, D is the distance, and v$_{Dop}$ is the measured Doppler shift of relevant tip. They adopted D = 1 kpc and $i = 16^\circ$.  If each finger pair has a different value of $i$, as we suspect, and if D = 420 pc, then their estimates of T$_{kin}$ need to be revised substantially.  Perhaps the best CO measure of the kinematic age of the lobes come from finger G1 at the tip of the N lobe  which, after a correction for D = 420pc, is roughly 500y.  In contrast, our estimate of 250y of the expansion age for the outer boundary of the lobes is based on proper motions for which D and $i$ are not needed.  

C00 did not attempt to derive the kinematic age for the dark dust lane.  However, the separations $\theta_r$ and Doppler velocities  v$_{Dop}$ are very similar for finger pairs A, B, C, and G in the dust lane, on the one hand, and D, E, and F in the lobes, on the other.  It is noteworthy that  the fingers, the lobes, and the dust lane may have similar ages in the range of 250 -- 500 y.

C00 also noted the presence of a small shell-like structure in close proximity to the nucleus whose radius is 1\arcsec.  The CO surface brightness drops precipitously within this shell.    We feel that the CO core could alternately be described as a $1\arcsec$ cavity at the center of the core of the CO brightness distribution. The radius of the hole corresponds to the volume within which the dust temperature drops from $\sim1000$K at its condensation radius of 26 msec to about 100K at $\approx$100 times further out (\S4).  

\subsection {The Kinematics of the Rings}

Without short interferometer spacings C00 could not detect the low-brightness, $\sim 40\arcsec$ halo outside of the core of CRL\,2688 in which all of the rings are situated.  However, \citet{2006ApJ...652.1626F} used the BIMA array in conjunction with the NRAO 12-m singe-dish telescope to map CRL\,2688 in  in the $^{12}$CO $J = 1Ð-0$ with $\sim3\arcsec$ resolution.  They detected CO from extending from an inner radius of $9\arcsec$ out to the noise level at almost $20\arcsec$.  Little velocity structure in this expanding region was reported.  This favors a constant-velocity outflow since uniform expansion would produce a large spread of Doppler shifts.  But the interferometer's poor sensitivity to very extended structure leaves a large margin of doubt in this conclusion.  

B01, C00, and Fong et al. noted the presence of the $^{12}$CO absorption feature that covers the bright core and the more extended halo.  As noted first byB01 the absorption arises in the extended  slow-velocity component.  Presumably this absorption region shares the same volume as the rings.  If so, then the rings lie in a foreground region of low velocity dispersion, $\pm 2$ km s$^{-1}$.  

The simplest interpretation of this result is that the entire ensemble rings results from repeated ejections, each of them characterized by outflow velocities blueshifted by $-18$ km s$^{-1}$.  The small CO absorption line width implies little variation in outflow speed over 4000y.  This result is inconsistent with the possibility of brief ejection and uniform expansion ($\dot\theta_r \propto \theta_r^1$) in \S3.3.

\subsection {The Distance to CRL\,2688}

If the ensemble of rings of CRL\,2688 expand isotropically, then their proper motion speed, $\dot\theta_r$ = $0\farcs07\pm0\farcs01$ in 6.65 y, is the same as the Doppler shift of the $^{12}$CO absorption feature from the nebular systemic velocity,  $\dot\theta_r$D = $18 \pm 2$ km s$^{-1}$. 
From this we derive D = $340 \pm 60$ pc, The uncertainty is dominated by the measurement of $\dot\theta_r$.  Alternately, if we adopt the width of the slow velocity component of B01, 15 km s$^{-1}$, then D = 280 pc.

\section {Conclusions}

Adopting a distance D = 420 pc and extinction A$_{\rm{V}}$ = 0.88 mag, then optical/IR continuum, CO, and H$_2$ images can be interpreted consistently as follows (in order of increasing distance from the nucleus): 
\begin{itemize}
\item The integrated SED and the optical-near IR color distribution is well-fitted with a realistic and generally tightly constrained  dust density distribution illuminated by a cool ($\sim$7000\,K) and luminous (5500 L$_\odot$) star.
\item The central shell (possibly a hole) in CO is associated with very warm dust at its condensation temperature needed to fit the nebular structure and the SED.
\item The lobes and dust lane are at least partially filled with finger-shaped outflows.
\item Inclination- and distance-independent expansion ages from the boundary of the lobes implies an expansion age $\approx$250y.  Based solely on CO kinematics, the fingers within the dark dust lane may have a similar outflow age.
\item The polar inclination angle lies between 5 and 15$^{\circ}$. We favor larger values in order to account for the lower brightness and redder colors of the south receding lobe. 
\item The system of rings has a kinematic signature that argues for repeated ejections at a constant, regulated velocity, 15 -- 18 km s$^{-1}$, along the line of sight.
\item Ring ejection continued for several thousand years preceding the formation of lobes.
\item A fit of the scattered starlight in the rings implies a density that decreases as radius$^3$.
\item As a result, the mass and momentum per ring have increased roughly linearly -- not unexpected as the luminosity and radius of the star increase during the ascent of the AGB. 
\item If the rings expand isotropically then we derive a distance of $340 pc \pm 60$ pc based on the measured proper motions of the rings and blueshift of the foreground $^{12}$CO  absorption feature.
\item At this distance the expansion time between rings -- that is their ejection period -- is 100y.
\item The entire nebula has faded irregularly by an average of $\approx10\%$ over the course of 6.65 y.  The northern parts of the nebula faded considerably more than the southern.  The changes are more likely the result of changes in the immediate environment of the star than the star itself.

\end{itemize}

\subsection{Evaluating Ring Formation Scenarios}

\citet{2008A&A...487..645R} has provided a thoughtful review of the mechanisms that drive AGB winds.  In addition, they successfully fitted models of constant AGB mass loss $\dot{\rm{M}}_{\rm{AGB}}$ to the CO profiles of many types of evolved outflow sources.  In contrast we find that $\dot{\rm{M}}_{\rm{AGB}}$ increases ($\sim$ linearly) in time. The same result was derived from a careful analysis of CO line profile measurements of other pPNe by \citet{2005ApJ...624..331H}.   Thus the fundamental issue of the evolution of $\dot{\rm{M}}_{\rm{AGB}}$ remains generally unresolved.

The process for forming the rings in the AGB winds is a matter of considerable discussion.  As noted by \citet{2001AJ....121..354B} and others, the rings are too closely spaced in time ($10^2-10^3$)y,  to be the result of thermal pulses since the typical interpulse time is $10^5-10^6$y (cf. \citealp{2010MNRAS.403.1413K, 2007PASA...24..103K} for M $\le$ 4 M$_\odot$ and Z = 0.02).  The periods of surface pulsations are far too short (Å1--10 y) to produce the arcs.  

The rings must form at the stellar surface or at the base of the flow since the opacity of the wind to driving radiation rapidly decreases with radius.  Their properties were predicted with the appropriate period using a two-fluid hydro model in which recurring large-scale dynamic instabilities alter the grain properties in the winds of a single AGB star \citep{2001A&A...371..205S}.   This concept was extended to two dimensions and applied to C stars by \citet{2006A&A...452..537W}.  Woitke argues that the extra degree of freedom generates local flow instabilities that produce multiple mushroom-shaped ejection events, the leading edges of which produce observable arcs.  \citet{2000ApJ...540..436S} proposed that the rings form as the result of periodic cycles of AGB magnetic activity that modulate the effective surface temperature of the star and, hence, the dust formation rate.

Alternately, the rings have been attributed to the periodic effects of a close companion star in an eccentric orbit with a small periastron and the appropriate period to explain the separations of the rings. \citet{1997ApJ...487..809H} introduced a class of models in which a companion star or planet  passes inside the grain formation region (very close to the star) so as to interrupt the grain formation and, hence, to modulate the density of the wind.  \citet{2007A&A...467.1081H} and \citet{2011apn5.confE.185R} proposed that if the close companion has an orbital eccentricity $e \le 0.8$ then a spiral wind outflow is extruded.  When the observer lies near the equatorial plane, as in CRL\,2688, the projected outflow pattern morphs to a series of round arcs on opposite sides of the that plane that alternate in radial displacement by even or odd multiples of the ring half-separation (that is, phased by 180$^\circ$). \footnote{ J. Cernicharo (private communication) has proposed a similar model for the rings of IRC+10216.}  Insofar as CRL\,2688 is concerned, the arcs are not sufficiently regular to test this hypothesis rigorously (Fig. \ref{polar}).

Finally, \citet{2002JAVSO..31....2Z} pointed out that the time scales of period changes in long-period "meandering Miras" ($\sim15\%$ of all Miras) and the inter-ejection times of rings are similar.  They propose that agreement in these times scales implies ``Mass loss [sic] variations induced by period changes may be one of the causes of the ringsÉ '' and they cite R Hya as an illustration.  (No physical mechanism that connects the phenomena is suggested in that paper.)  While the exposed surface of the central star of CRL/,2688 is presently too hot to be a long-period Mira, it may have been as recently as a few centuries ago when the lobes were ejected.

\begin{acknowledgments}
A special thanks to Max Mutchler of STScI for help with the image calibrations from different HST cameras.  The anonymous referee of this paper made very helpful corrections and thoughtful suggestions.

This project was supported by HST GO grant 11580. Support for GO11580 was provided by NASA through a grant from the Space Telescope Science Institute, which is operated by the Association of Universities for Research in Astronomy, Incorporated, under NASA contract NAS5-26555. TG is grateful for support from a Boeing Scholarship and a McNair Fellowship awarded through the office of Undergraduate Academic Affairs at the University of Washington.  JA is  partially supported by the Spanish MICINN, program CONSOLIDER INGENIO 2010, grant ``ASTROMOL'' (CSD2009-00038).  Numerical modeling by DV was performed on computer cluster Hybrid (\url{www.gpuhybrid.org}) financed by the National Foundation for Science, Higher Education and Technological Development of the Republic of Croatia.

Some of the data presented in this paper were obtained from the Multimission Archive (MAST) at the Space Telescope Science Institute (STScI).  STScI is operated by the Association of Universities for Research in Astronomy, Inc., under NASA contract NAS5-26555. Support for  MAST for non-HST data is provided by the NASA Office of Space Science via grant NAG5-7584 and by other grants and contracts.
\end{acknowledgments}

{\it Facility:} \facility{HST (WFC3)}

\clearpage\pagebreak
\appendix
\section{Appendix}

We describe the details of the major structural components used in or derived from the model.  The absorption and scattering properties of the dust shown in Fig. \ref{DejanB} were adopted for the model, as discussed for each component of the model below.

{\it Star:} The central star is not visible in images due to large dust opacity ($\approx55$) along the line of sight. Nonetheless, its basic spectral type can be derived from the molecular lines in the spectrum measured in the scattered light from bipolar lobes. Using this approach, \cite{1975ApJ...198L.135C} classified the stellar spectra as of type F5 Ia, while \cite{1977PASP...89..829C} argue that it is F2 I. Thus we initially fixed the stellar temperature as a 7000K black-body spectrum which is similar to the temperatures suggested by these spectral types.

In the end we found that several combinations of stellar spectral types and adopted reddening properties of the polar caps provide good fits to the nebular colors and the SED provided that we fix the luminosity at 5500 L$_\sun$.  For example, a very satisfactory fit is obtained if we fix T$_*$ at 7000K and impose internal reddening consistent with $\sim2$ magnitudes of extinction in the dense polar caps. However, we were also able to create a comparably successful model with a 4000K black-body star of about the same luminosity and no internal reddening.  If we adopt the early-F-type spectral type of about L$_* \sim5500$ L$_\sun$ and the necessary internal reddening then the spectral type of the star is between F1-Ib and F3-Ib.  (If the distance is doubled then the spectral type becomes F1-Iab and F3-Iab.)

{\it Central dust free cavity:}  F-type stars are far too hot to form dust in their atmospheres.  So we assumed that the innermost dust is located at the perimeter of a (presumed) spherical cavity of radius $R_{in}$ in units of stellar radii.  The value of $R_{in}$ is determined in part by the star's luminosity L${_*}$, where L${_*} = 5500 \pm 20\%$ is measured by fitting the SED of the composite nebula at the adopted distance of 420 pc.  $R_{in}$ is basically set by the requirement that the temperature along the walls of the cavity, T$_{\rm{max}}$, is less than condensation temperature of the dust T$_{\rm{cond}}$, although T$_{\rm{cond}}$ is not known unless the chemical nature of the dust is somehow determined.  Unfortunately, the high disk opacity precludes near-IR measurements of the color temperature of the cavity which might at least constrain the adopted value of T$_{\rm{max}}$.  For the same reasons the size and geometry of the cavity cannot be measured.  Moreover, we found that suitable models could be developed over a substantial range of adopted values of T$_{\rm{max}}$.  Ultimately we simply adopted T$_{\rm{max}}$ = 1000 K in the midplane.  It then follows that $R_{in} = 46.4R_*  $= 0.56 msec with a formal error of 10\% based on luminosity uncertainties.  (Given the many assumptions, the actual uncertainty is obviously larger.)

{\it Disk:} Imaging reveals a dense, flattened, and modestly articulated dusty structure in the midplane of CRL\,2688. Its irregular shape is exposed by multiply scattered starlight from the lobes (see Fig. \ref{Color}) and supported by the morphologies of H$_2$ and radio molecular images noted earlier.  For modeling purposes we call this structure an oblate ``disk''.  The SED slope between 0.35 and 3.6 cm points toward emission from large dust grains, with sizes of 0.5 cm or more (\cite{2000ApJ...528L.105J}). On the other hand, the 3.8 and 4.7$\micron$ images by G02 showed that the dust grains are not larger than $\sim0.2\micron$. 

This mismatch is a result of inhomogeneous spatial distribution of grain sizes, where large grains probably exist closer to the star in the irregular core of the dusty structure, while smaller grains dominate the opacity in the outer edges of the disk.  We simplify the disk structure using single-grain dust of size $0.4\micron$ with a radial density distribution that varies as $r^a$, where $r$ is the distance form the star and $a$ is a free parameter. We adopted an oblate ellipsoidal disk shape with bipolar lobes carved out at polar angles $\la15^\circ$. For simplicity we assume that the edges of the bipolar cones are abrupt, so that the shadow that they cast in the envelope is sharply defined. The semi-major axis of the disk is $\sim7\arcsec$ and the semi-minor axis is $\sim1\farcs8$. The disk's midplane optical depth is adequate to hide the star at $\lambda \la 4.5\micron$.

{\it Polar caps:} Polar caps are a vital construct of the model that collimates the light that both illuminates the lobes and escapes to form the searchlights.  From the Earth the line of sight to the caps traverses the dense disk, so the caps are not fully visible directly in HST images in F606W and F814W. (Parts of the caps may be seen through patchy extinction in the F110W and F160W images.)

We noticed that without the caps, the scattering model for illumination from a central star of T$_* \approx $ 7000 K produces lobes that are too bright and too blue to fit the data.  Thus the starlight is modulated by an unobserved and compact region of dense dust.  The only way  to fit the shape of spectral continuum in the visible range with a 7000K star is to redden and partially absorb the light that enters the lobes to match the illumination expected from an illumination source $\ga$4000K .  In essence, the caps are dust density enhancements at the base of lobes and, so, may be parts of a single entity.  However, the requisite dust grains differ in size from the larger ones, $0.4\micron$ used in the lobes and the disk.  The dust in the caps consists of particles from from 0.01 to $0.05\micron$ that follow a radius$^{-3.5}$ size distribution as shown in the lower curve of Fig. \ref{DejanB}. Without observations the density of these grains is difficult to characterize. We adopted a constant density $\rho_{\rm{cap}}$ within a cone segment where the cone height R$_{\rm{cap}}$ and the opening angle $\theta_{\rm{cap}}$ are free parameters. This high-density cone is added to the densities of the lobe and disk present in this region. 

{\it Lobes:} The bipolar lobes are the brightest of all features detected in our images of CRL\,2688. The (identical) lobes have sharp edges in HST images, which enables precise tracing of their geometry for distances larger than about $1\farcs5$ from the position of the central star.  Our model is symmetric with respect to the midplane. We analytically describe the shape of each lobe as an axially truncated cone with apex at the nucleus plus half of an ellipsoid of semi-major axis = $7\arcsec$ attached to the base of the cone (Fig. \ref{DejanA}) They join where their mutual widths are $1\farcs26$. The cones have a total opening angle of $30^\circ$.  The internal densities follow a power law, $r^b$ where b is varied to reproduce the observed radial trends in lobe surface brightness.   We adopted an isotropic dust scattering function.  Dust grain size in lobes must be dominated by grains larger than about $0.1\micron$ to assure that the lobes remain bright as seen in near infrared images. For simplicity we adopted the same dust properties as in the disk.

Although the lobes are directly illuminated by the central star, their surface brightnesses are almost flat owing to multiple interior scatterings.  However, the northern lobe is observed to be slightly brighter then the southern lobe.  The only way how we can achieve this asymmetry is through an axial inclination of angle $i$ and the assumption of internal reddening within the envelope. A value of $i \sim 15^\circ$ provides a good match to the lobe ratio for the properties of the envelope that we assumed.  Smaller values of $i$ provide slightly better fits to the projected appearance of the arcs illuminated within the searchlights.

Based on their HST imaging, S98b proposed a model for searchlight beams where the starlight initially escapes through a clear hole on the polar axis near the nucleus.  A shadow zone within the searchlight is the result of shadows cast by a pair of extended knots lying at the outer tips of the the lobes. The existence of the knots is motivated by features seen at the tips of the lobes seen in H$_2$ and CO images.  In principle, the shadow-casting knots can be located at any radius along the polar axis. We created an annular aperture by locating a cone of enhanced density within each lobe whose opening angle, $10^\circ$ in total, to fit the shadow edges within the searchlights.  The dust within this inner cone is like that in the disk and outer lobes.  Its density varies as $r^c$ where $c$ is determined by fitting the darkness of the shadows.

{\it Envelope:} CRL\,2688 shows a faint spherical nebulosity out to angular radii $r \sim 40\arcsec$ from the star. We call this region the nebular ``envelope'' within which thin spherical shells mimic the conspicuous co-centric rings or arcs of varying size and brightness. For simplicity and speed we fixed the envelope radius to be twice the size of the lobes, $\sim 15\arcsec$. This is big enough to analyze the properties of the rings and to collect all the flux at mid-IR wavelengths where the SED reaches its maximum and begins to decline.

We use dust of uniform radius in the envelope.  The grain radius is a free parameter.  The envelope density profile is constrained by its measured brightness in scattered light. To match the general properties of the observed envelope we assume that the dust density distribution $\propto r^{-3}$ (i.e., a mass loss rate that increases linearly with time for constant outflow velocity, a common feature among pPNe [\citealp{2005ApJ...624..331H}]). This produces a brightness profile $\sim r^{-4}$, similar to the measured trend of $\sim r^{-3.7}$ found by S98b. The composite image on the right side of Fig. \ref{DejanA} shows a bluish color of envelope compared to the disk color, indicating that the envelope contains smaller grains than the disk and the lobes. However, the color of the searchlights is blue only in proximity to the blue lobes that illuminate them locally after multiple scattering.  The color reddens at larger radii where starlight emerging from the caps is the primary light illumination source.  This trend is not observed in our WFC3 images, perhaps because the grains in the envelope contain many smaller particles than those we used in the model.

{\it Shells:} S98b concluded that the observed rings (represented by spherical shells in our model) are local density enhancements. They argued that a density enhancement by a factor of three yields a brightness increase of a factor of two. We used these density enhancement values relative to the ambient density of the background envelope that declines as $r^{-3}$. Since observed arcs are broken and deformed, our aim is not to reproduce them in detail and we focus only to their general brightness properties.

{\it Spectral energy distribution:} matching the SED is one of the greatest challenges---and successes---of our model.  We fitted the SED after assuming that interstellar dust along the line of sight reddens the visible and near-IR light, as described in \S4.  However, we note that we also managed to derive successful models of the SED without interstellar reddening, but this is merely of academic interest. We also plot centimeter fluxes from \cite{2000ApJ...528L.105J} even though our model does not address the emission at radio wavelengths.  Presumably the radio $\lambda$mm and sub-mm flux is the direct result of thermal dust emission which is outside the scope of our model.

{\it Technical Details of the Model.}  The fitted model parameters are listed in Table 4. Some do not have units.  The reason is that  the model works with dimensionless variables, where the scale is set by the optical depth. Thus we do not parameterize the mass and density of dust particles in physical units. The absolute scale is set by the total radial optical depth in the midplane.  A detailed discussion on this modeling issue is given in \citep{2003PhDT........40V}.

The dust densities are not explicitly specified in the model because the absolute density values are scaled out in radiative transfer equations and replaced with scattering and absorption optical depths. The calculation determines the optical depth along an arbitrary radial line and an arbitrary wavelength. We use the optical depth $\tau_{\rm{0.55}}$ in the midplane at $0.55\micron$.  The overall algorithm describing dust density at a point $(x,y,z)$, for a given semi-major axis of disk, R$_{\rm{disk}}$ (semi-minor axis = 0.25 R$_{\rm{disk}}$) and lobe R$_{\rm{lobe}}$ (semi-minor axis = 0.18 R$_{\rm{lobe}}$), can be symbolically described as this:

\indent
$r$=x$^2$+y$^2$+z$^2$\\
\indent
if ($r$ $<$ 1) then density=0\\
\indent
else if (x $<$ z $\cdot$ tan($\theta_{lobe}$/2) AND x$^2 <$ 0.18$^2$ $\cdot$ ($R_{lobe}^2$ - z$^2$)) then $\{$\\
\indent \indent
if (x $<$ z $\cdot$ tan($\theta_{shadow}$/2) then density $\propto r^c$\\
\indent \indent
else density $\propto r^c$\\
\indent
$\}$ else if (z$^2 <$ 0.25$^2 \cdot$ ($R_{disk}^2$ - x$^2$)) then  density $\propto r^a$\\
\indent
else $\{$\\
\indent \indent
  density $\propto r^{-3}$\\
\indent \indent
  if ($r$ within shell) then density $\equiv 3 \cdot$ density\\
\indent
$\}$\\
\indent
if ($r < R_{cap}$ AND x $<$ z $\cdot$ tan($\theta_{cap}$/2)) density $\equiv$ density + $\varrho_{cap}$\\

\clearpage\pagebreak
\begin{table}[!th]
\begin{center}
\caption{Log of HST/WFC3 Images of CRL\,2688 from G011580}\label{T1}
\begin{tabular}{l c c c r l }
\hline\hline
WFC3  & ~ & Filter  center/& Pixel  & Exposure &  Archive \\
Camera  & Filter & width ($\mu\rm{m}$) & Size (\arcsec) & Time (s) & Data Set \\
\hline
UVIS1/SUB512 & F606W & 0.59/0.22 & 0.040 & 100 & IB1M01011\\
UVIS1/SUB512 & F606W & 0.59/0.22 & 0.040 & 600 & IB1M01021\\
UVIS1/SUB512 & F814W & 0.80/0.15 & 0.040 & 150 & IB1M01031\\
UVIS1/SUB512 & F814W & 0.80/0.15 & 0.040 & 1000 & IB1M01041\\
IR/SUB512 & F110W & 1.15/0.44 & 0.128 & 70 & IB1M05RMQ\\
IR/SUB512 & F110W & 1.15/0.44 & 0.128 & 1178 & IB1M05020\\
IR/SUB512 & F160W & 1.54/0.27 & 0.128 & 47 & IB1M05RNQ\\
IR/SUB512 & F160W & 1.54/0.27 & 0.128 & 1178 & IB1M05010\\
\hline
\end{tabular}
\end{center}
\end{table}

\clearpage\pagebreak
\begin{table}
\begin{center}
\caption{Log of HST F606W Images of CRL\,2688}\label{T2}
\begin{tabular} {l c c c r r}
\hline\hline
~ & Pixel  & Exposure & Observation& GO & Archive ~~\\
Camera & Size (\arcsec) & Time (s) & Date & Program & Data Set ~~\\
\hline
WFPC2/WF & 0.10 & 260 & 07/17/95 & 6221 & U2RC0303T\\
WFPC2/WF & 0.10 & 60  & 04/09/98 & 7423 & U4SHA101R\\
WFPC2/WF & 0.10 & 200 & 04/09/98 & 7423 & U4SHA102R\\
WFPC2/WF & 0.10 & 200 & 04/09/98 & 7423 & U4SHA103R\\
WFPC2/WF & 0.10 & 200 & 04/09/98 & 7423 & U4SHA104R\\
WFPC2/WF & 0.10 & 200 & 04/09/98 & 7423 & U4SHA105R\\
ACS/WFC & 0.050 & 110 & 10/16/02 & 9586 & J8GH55PLQ\\
ACS/WFC & 0.050 & 360 & 10/16/02 & 9586 & J8GH55PNQ\\
ACS/HRC & 0.025 & 200 & 12/10/02 & 9463 & J8DI81011\\
ACS/HRC & 0.025 & 200 & 12/10/02 & 9463 & J8DI81041\\
ACS/HRC & 0.025 & 200 & 12/10/02 & 9463 & J8DI81051\\
WFC3/UVIS & 0.040 & 100 & 08/05/09 & 11580 & IB1M01011\\
WFC3/UVIS & 0.040 & 600 & 08/05/09 & 11580 & IB1M01021\\
\hline
\end{tabular}
\end{center}
\end{table}

\clearpage\pagebreak
\begin{table}
\begin{center}
\caption{Methods \& Results for Arcs \& Rings}\label{T3}
\begin{tabular}{l c c c r }
\hline\hline
Cameras & Interval & Method & $M$ & Exp. Age (y) \\
\hline
WFPC2-WF3/WFC3 & 2009.59 - 1995.54 & Difference & 0.013$\pm$0.003 & 1076$\pm$30\% \\
WFPC2-WF3/WFC3 & 2009.59 - 1998.27  & Difference &  0.013$\pm$0.003 & 871$\pm$35\% \\
WFPC2-WF3/WFC3 & 2009.59 - 1998.27 & Ratio & 0.012$\pm$0.003 & 943$\pm$35\%\\ACS-HRC/WFC3 & 2009.59 - 2002.94 & Difference & 0.010$\pm$0.002 & 665$\pm$20\% \\
ACS-HRC/WFC3 & 2009.59 - 2002.94  & Ratio & 0.009$\pm$0.002 & 739$\pm$10\%\\
\hline
\end{tabular}
\end{center}
\end{table}

\clearpage\pagebreak
\begin{deluxetable}{lccl}
\centering
\tabletypesize{\scriptsize}
\tablecaption{\label{table-fit} List of model parameters
}
\tablewidth{0pt}
\tablehead{
\colhead{Parameter description} &
\colhead{Parameter} &
\colhead{Value} &
\colhead{Comment}
}

\startdata
lobe optical depth &  $\tau_{lobes}$ & 2 &
required to explain the lobe colors\\
along polar axis
\\\hline

disk optical depth &  $\tau_{disk}$ & 55 &
it has to be large to hide the star in the near IR; additional\\
in the midplane &&&  constraint comes from the strength and spectral shape of\\
 &&&  mid IR flux; variations within 10\% are acceptable \\\hline 

outer grid radius & $R_{out}$ & $600R_{in}$ & arbitrarily fixed; the detected nebulosity extends further out \\\hline 

stellar temperature	& $T_{*}$ & 7000K &	black body spectrum \\\hline 

maximum dust & $T_{max}$ & 1000K & poorly constrained; models with larger or smaller $T_{max}$ are \\
temperature in & & & allowed; this is maximum temperature of dust grains in \\
the midplane & & &  the midplane \\\hline 

disk density profile & $r^{a}$ & $a= - 0.5$ & poorly constrained; variations $\pm$100\% are allowed \\\hline 

disk semi-major axis & $R_{disk}$ & $300R_{in}$	& this value of $R_{disk}$ corresponds to 7-8$\arcsec$ \\
 & & & (based on the luminosity derived from the SED fit)\\\hline 

disk, lobe and shadow &  & $0.4{\mu}m$ & see Figure \ref{DejanB}; poorly constrained; \\
dust grain radius & & \\\hline 

lobe density profile & $Br^b$ & $B=2\times 10^{-5}$ & constrained by lobes brightness; $B$ has to be small enough \\
                     &        & $b=1$ &  to make lobes optically thin\\\hline 

lobe semi-major axis & $R_{lobe}$ & $300R_{in}$ & based on luminosity derived from the SED fit, this value \\
& & & of $R_{lobe}$ corresponds to 7-8$\arcsec$ \\\hline 

lobe opening angle &	$\theta_{lobe}$ & $30\degr$	& constrained by lobe images\\\hline 
shadow opening angle &	$\theta_{shadow}$ & $10\degr$	& constrained by ''searchlights'' images\\\hline 

shadow density & $Cr^c$ & $C=2000$ & poorly constrained, but its optical depth has to be larger \\
profile & & $c=-3$ & than 1; steep density profile puts the most of optical \\
 & & & depth in the base of lobes\\\hline 

polar cap density & $\varrho_{cap}$ & 6500 & equivalent to the optical depth of $\sim$2 at 0.55${\mu}m$, which \\
 & & & dims and reddens the starlight entering the lobes\\\hline 

polar cap radius & $R_{cap}$ & $2 R_{in}$ & poorly constrained; optical depth of the cap is the key \\
& & & parameter\\\hline 

polar cap opening &	$\theta_{cap}$ & $40\degr$	& poorly constrained, but it must be $>\theta_{lobe}$ lobe opening \\
angle & & & angle;  optical depth of the cap is the key parameter\\\hline 

envelope density  &	$Dr^{-3}$ & $D=4.2\times 10^4$ & constrained by the envelope brightness profile \\
profile & & & \\\hline 

envelope dust & & $0.2{\mu}m$ & see Figure \ref{DejanB}; poorly constrained; must be large enough \\
grain radius & & & to keep envelope brightness high in the near IR, \\
 & & & but smaller than grains in the disk\\\hline 

geometrical thickness &	 $\Delta r_{shell}$ & $15 R_{in} (0\farcs39)$ & poorly constrained because arcs are of irregular shape\\
of shells & & & and thickness\\\hline 

radial distance & $d r_{shell}$ & $35 R_{in} (0\farcs9)$  & poorly constrained because arcs are of irregular shape\\
between shells & & & and thickness \\\hline\hline 

\multicolumn{4}{c}{Derived parameters}\\\hline 

inclination angle & $i$ & $15\degr$ & constrained by the brightness asymmetry between \\
 & & & the northern and southern lobe\\\hline 

bolometric flux & $F_{bol}$ & $1.0\pm 0.2\times 10^{-9}$ & fit to the data gives 12\% smaller flux, while this value \\
($W/m^2$) & & & takes into consideration the asymmetry in diffuse radiation\\\hline 

luminosity ($L_\odot$) & $L_*$ & $5500\pm 1100$ & based on 420pc distance and SED fit \\\hline 

stellar radius ($R_\odot$) & $R_*$ & $51\pm 5 R_*$ & based on luminosity, stellar temperature \\
 & & $0.56\pm 0.05$ msec & and distance = 420pc \\\hline 

dust condensation & $R_{in}$ & $46.4 R_*  (0\farcs026)$ & critically depends on $T_{max}$ ,after that on $\tau_{0.55}$\\
radius & & &        and optical properties of dust in disk\\

\enddata
\end{deluxetable}

\bibliographystyle{apj}
\bibliography{balick}

\end{document}